\documentclass[11pt]{article}

\usepackage[margin=1.1in]{geometry}

\usepackage[cmex10]{amsmath}
\usepackage[utf8x]{inputenc}
\usepackage[nocompress]{cite}
\usepackage{graphicx, multirow, booktabs, color, listings}
\usepackage{balance}

\newcommand{\x}{{\bf x}}
\newcommand{\y}{{\bf y}}

\usepackage{amssymb, amsmath, amsthm}
\usepackage{thmtools}

\usepackage{mathtools}

\begin{document}

\title{Multiword Arithmetic and Parallel Computing}

\author{Jan Verschelde\thanks{Supported by a 2023 Simons Travel Award.
University of Illinois at Chicago,
Department of Mathematics, Statistics, and Computer Science,
851 S. Morgan St. (m/c 249), Chicago, IL 60607-7045,
Email: {\tt janv@uic.edu}, URL: {\tt http://www.math.uic.edu/$\sim$jan}.}
}

\date{26 October 2025}


\maketitle

\begin{abstract}
In many applications, the precision by the available hardware arithmetic
is insufficient to guarantee accurate results.
Multiword arithmetic is a special type of multiprecision arithmetic
where a multiple double is an unevaluated sum of 64-bit doubles, or
where a multiple integer is an unevaluated sum of 64-bit integers.
Parallel computing is applied to compensate for the cost overhead
of multiword arithmetic.
This type of arithmetic exploits naturally the optimized hardware,
allows for efficient type conversions, memory layouts, all favorable 
for parallel computing.  For example, storing a multiword in registers 
rather than arrays is beneficial to parallel computing by tasking
and acceleration by graphics processing units.
Code for multiword arithmetic is available in the software PHCpack,
written mainly in Ada,
publicly available at github, and as an Alire crate,
released under the GNU GPL v3.0 license.
\end{abstract}


\section{Introduction}

An algorithm is {\em robust} if it does not fail for small perturbations
of degenerate inputs~\cite{She97}.
Software can be made more robust by the application
of multiprecision arithmetic.  
The algorithms~\cite{MBDJJLMRT18}
to extend 32-bit floating-point arithmetic using
originated in the late sixties~\cite{Dek71} and available software
packages are QDlib~\cite{HLB01} and CAMPARY~\cite{JMPT16}.

A {\em multiple double} is an unevaluated sum of nonoverlapping doubles.
The multiple double arithmetic multiplies the accuracy of the results,
from 16 decimal places to 32 and 64, respectively for double doubles 
and quad doubles.  However, the cost overhead is significant, 
and parallel computing allows to compensate for this cost overhead.
The overhead factors are summarized in Table~\ref{tabOverhead}.

\begin{table}[hbt]
\begin{center}
\begin{tabular}{r|rrr|r}
  & \multicolumn{1}{c}{\tt add}
  & \multicolumn{1}{c}{\tt mul}
  & \multicolumn{1}{c|}{\tt div}
  & \multicolumn{1}{c}{\tt avg} \\ \hline
 2 &   20  &    23  &    70 &   37.7 \\
 4 &   89  &   336  &   893 &  439.3 \\
 8 &  269  &  1742  &  5126 &  2379.0 \\
16 &  925  & 11499  & 33041 & 15155.0 \\
\end{tabular}
\caption{The number of operations with doubles for a multiple
double addition ({\tt add}), multiplication ({\tt mul}),
and division ({\tt div}), required for $m$-double arithmetic,
for $m = 2, 4, 8, 16$.}
\label{tabOverhead}
\end{center}
\end{table}

This paper describes some recent additions to PHCpack~\cite{Ver99},
extending contributions described in~\cite{Ver20} and~\cite{Ver22}.
One application is the computation of power series~\cite{Ver24}.
As the errors in the leading coefficients propagate to the trailing
coefficients, the leading coefficients must be computed at higher
accuracy than what can be computed in 64-bit double precision.

The code described in this paper serves as a computational preparation
for use of tensor cores on a graphic processing unit,
in a stepping stone to implement the Ozaki scheme~\cite{OOOR12}.

\section{An Error Free Summation}

\begin{figure}[hbt]
\begin{center}
\begin{picture}(245,90)(-5,0)
\put(-5,0){$=$}
\put(-5,80){$+$}
\put(-5,60){$+$}
\put(-5,20){$+$}
\put(5,0){\line(1,0){40}} \put(5,0){\line(0,1){6}}
\put(5,6){\line(1,0){40}} \put(45,0){\line(0,1){6}}
\put(65,0){\line(1,0){40}} \put(65,0){\line(0,1){6}}
\put(65,6){\line(1,0){40}} \put(105,0){\line(0,1){6}}
\put(85,-4){\line(0,1){12}}
\put(123,0){$=$} \put(181,0){$=$}
\put(120,3){\vector(-1,0){10}}
\put(60,3){\vector(-1,0){10}}
\put(133,0){\line(1,0){20}} \put(133,0){\line(0,1){6}}
\put(133,6){\line(1,0){20}} \put(153,0){\line(0,1){6}}
\put(155,1){\tiny $b\!\cdots\!b$}
\put(153,0){\line(1,0){20}}
\put(153,6){\line(1,0){20}} \put(173,0){\line(0,1){6}}
\put(191,0){\line(1,0){20}} \put(191,0){\line(0,1){6}}
\put(191,6){\line(1,0){20}} \put(211,0){\line(0,1){6}}
\put(213,1){\tiny $b\!\cdots\!b$}
\put(211,0){\line(1,0){20}}
\put(211,6){\line(1,0){20}} \put(231,0){\line(0,1){6}}
\put(-5,12){\line(1,0){50}}
\put(123,12){\line(1,0){50}}
\put(181,12){\line(1,0){50}}
\put(5,80){\line(1,0){40}} \put(5,80){\line(0,1){6}}
\put(5,86){\line(1,0){40}} \put(45,80){\line(0,1){6}}
\put(50,83){\vector(1,0){10}}
\put(65,80){\line(1,0){40}} \put(65,80){\line(0,1){6}}
\put(65,86){\line(1,0){40}} \put(105,80){\line(0,1){6}}
\put(85,76){\line(0,1){14}}
\put(110,83){\vector(1,0){15}}
\put(133,80){\line(1,0){20}} \put(133,80){\line(0,1){6}}
\put(133,86){\line(1,0){20}} \put(153,80){\line(0,1){6}}
\put(155,81){\tiny $0\!\cdots\!0$}
\put(153,80){\line(1,0){20}}
\put(153,86){\line(1,0){20}} \put(173,80){\line(0,1){6}}
\put(191,80){\line(1,0){20}} \put(191,80){\line(0,1){6}}
\put(191,86){\line(1,0){20}} \put(211,80){\line(0,1){6}}
\put(213,81){\tiny $0\!\cdots\!0$}
\put(211,80){\line(1,0){20}}
\put(211,86){\line(1,0){20}} \put(231,80){\line(0,1){6}}
\put(5,60){\line(1,0){40}} \put(5,60){\line(0,1){6}}
\put(5,66){\line(1,0){40}} \put(45,60){\line(0,1){6}}
\put(50,63){\vector(1,0){10}}
\put(65,60){\line(1,0){40}} \put(65,60){\line(0,1){6}}
\put(65,66){\line(1,0){40}} \put(105,60){\line(0,1){6}}
\put(85,56){\line(0,1){14}}
\put(110,63){\vector(1,0){10}}
\put(123,60){$+$} \put(181,60){$+$}
\put(133,60){\line(1,0){20}} \put(133,60){\line(0,1){6}}
\put(133,66){\line(1,0){20}} \put(153,60){\line(0,1){6}}
\put(155,61){\tiny $0\!\cdots\!0$}
\put(153,60){\line(1,0){20}}
\put(153,66){\line(1,0){20}} \put(173,60){\line(0,1){6}}
\put(191,60){\line(1,0){20}} \put(191,60){\line(0,1){6}}
\put(191,66){\line(1,0){20}} \put(211,60){\line(0,1){6}}
\put(213,61){\tiny $0\!\cdots\!0$}
\put(211,60){\line(1,0){20}}
\put(211,66){\line(1,0){20}} \put(231,60){\line(0,1){6}}
\put(25,40){$\vdots$}
\put(83,40){$\vdots$}
\put(151,40){$\vdots$}
\put(209,40){$\vdots$}
\put(5,20){\line(1,0){40}} \put(5,20){\line(0,1){6}}
\put(5,26){\line(1,0){40}} \put(45,20){\line(0,1){6}}
\put(50,23){\vector(1,0){10}}
\put(65,20){\line(1,0){40}} \put(65,20){\line(0,1){6}}
\put(65,26){\line(1,0){40}} \put(105,20){\line(0,1){6}}
\put(85,16){\line(0,1){14}}
\put(110,23){\vector(1,0){10}}
\put(123,20){$+$} \put(181,20){$+$}
\put(133,20){\line(1,0){20}} \put(133,20){\line(0,1){6}}
\put(133,26){\line(1,0){20}} \put(153,20){\line(0,1){6}}
\put(155,21){\tiny $0\!\cdots\!0$}
\put(153,20){\line(1,0){20}}
\put(153,26){\line(1,0){20}} \put(173,20){\line(0,1){6}}
\put(191,20){\line(1,0){20}} \put(191,20){\line(0,1){6}}
\put(191,26){\line(1,0){20}} \put(211,20){\line(0,1){6}}
\put(213,21){\tiny $0\!\cdots\!0$}
\put(211,20){\line(1,0){20}}
\put(211,26){\line(1,0){20}} \put(231,20){\line(0,1){6}}
\end{picture}
\caption{Split a vector of doubles, add the parts, fuse the result.}
\label{figErrorFreeSum}
\end{center}
\end{figure}
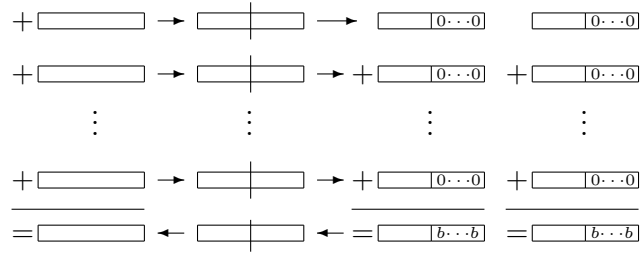

Assuming all 64-bit doubles have the same exponent,
we work with 52-bit integers (fractions of the doubles).
The idea for an error free summation is introduced
in Figure~\ref{figErrorFreeSum}.

If the number of additions does not exceed some threshold,
then we have sufficiently many zero bits left at the end of the numbers
to represent the result exactly, without any error.

The idea in Figure~\ref{figErrorFreeSum} will be applied
to computing inner products with double double arithmetic.
Given are vectors $\x$ and $\y$ both of length $n$,
of double double numbers, 
we compute $\displaystyle \sum_{k=1}^n x_k \star y_k$,
where $\star$ is the double double multiplication.

The double double $x_k$ is represented by 
$(x^{\mbox{hi}}_k, x^{\mbox{lo}}_k)$, where 
the high double $x^{\mbox{hi}}_k$ and
the low double $x^{\mbox{lo}}_k$ of $x_k$ are splitted in quarters:

\begin{displaymath}
 (\overbracket[1.2pt][6pt]{x_{k,0},x_{k,1},x_{k,2},x_{k,3}}^{x^{\mbox{hi}}_k},
  \overbracket[1.2pt][6pt]{x_{k,4},x_{k,5},x_{k,6},x_{k,7}}^{x^{\mbox{lo}}_k}).
\end{displaymath}
After splitting also $y_k$, we compute in double arithmetic:
\begin{displaymath}
   s_0 = \sum_{k=1}^n x_{k,0} y_{k,0}, ~
   s_1 = \sum_{k=1}^n x_{k,1} y_{k,0} + x_{k,0} y_{k,1},
\end{displaymath}
and
\begin{displaymath}
   s_i = \sum_{k=1}^n \sum_{j=0}^i x_{k,j} y_{k,i-j},
\end{displaymath}
for $i=2,\ldots,7$.  Then,
add $s_0 + s_1 + \cdots + s_7$ in double double arithmetic.

To examine the computational efficiency, random 64-bit doubles
are generated with a fraction of 52 bits in following pattern:

\begin{displaymath}
   1 \underbrace{b b \cdots b}_{\mbox{12 bits}}
   1 \underbrace{b b \cdots b}_{\mbox{12 bits}}
   1 \underbrace{b b \cdots b}_{\mbox{12 bits}}
   1 \underbrace{b b \cdots b}_{\mbox{12 bits}}, \quad b \in \{0, 1\}.
\end{displaymath}

Splitting such double into four leads to doubles with fractions

\begin{displaymath}
   \begin{array}{c}
      1 b \cdots b ~ 0 0 \cdots 0 ~ 0 0 \cdots 0 ~ 0 0 \cdots 0, \\
      0 0 \cdots 0 ~ 1 b \cdots b ~ 0 0 \cdots 0 ~ 0 0 \cdots 0, \\
      0 0 \cdots 0 ~ 0 0 \cdots 0 ~ 1 b \cdots b ~ 0 0 \cdots 0, \\
      0 0 \cdots 0 ~ 0 0 \cdots 0 ~ 0 0 \cdots 0 ~ 1 b \cdots b.
   \end{array}
\end{displaymath}

By virtue of the placement of the ones in the random fractions,
all quarters have fixed exponents, e.g.: 0, $-13$, $-26$, $-39$.
All doubles in a multiple double are generated according this pattern.

\section{Computational Results}

The results on
computing 1,024 times $\displaystyle \sum_{k=1}^{6144} a_k \star b_k$
in increasing precision is shown in Table~\ref{tabResults}.

\begin{table}[hbt]
\begin{center}
\begin{tabular}{r|rc|c|rc}
    & \multicolumn{2}{c|}{ordinary}
    & {\small \textcolor{red}{\bf speedup}}
    & \multicolumn{2}{c}{vectorized} \\ \hline
16d & 40s 780ms & 6.3x & \textcolor{red}{\bf 4.3x} & ~9s 491ms & 6.2x \\
 8d &  6s 428ms & 3.3x & \textcolor{red}{\bf 4.2x} &  1s 520ms & 4.8x \\
 4d &  1s 977ms & 12.x & \textcolor{red}{\bf 6.2x} &     318ms & 4.6x \\
 2d &     158ms & 13.x & \textcolor{red}{\bf 2.3x} &      69ms & 2.3x \\
 1d &      12ms &      & \textcolor{red}{\bf 0.4x} &      30ms &
\end{tabular}
\caption{Rows for double (1d), double double (2d), quad double (4d),
octo double (8d), and hexa double (16d), comparing the new vectorized
inner products to the ordinary ones.
Times are in milliseconds (ms).
The column next to the times are the cost overhead factors. }
\label{tabResults}
\end{center}
\end{table}

Times in Table~\ref{tabResults} were obtained 
on an Intel Xeon 5318Y Ice Lake-SP, up to 3.40GHz,
256GB of internal memory at 3200MHz, GNU/Linux, Microway 2024,
compiled with GNAT 12.2.0, flags {\tt -O3 -gnatp -gnatf}.

It takes 9 seconds for 1,024 inner products in hexa double precision (16d).
The wall clock time is 9s 308ms, with 85ms for generating the vectors.

In a high level multithread computation, 
every thread does one inner product.
On two 24-core Intel Xeon 5318Y Ice Lake-SP, up to 3.40GHz,
256GB of internal memory at 3200MHz, GNU/Linux, Microway 2024,
 compiled with GNAT 12.2.0, flags {\tt -O3 -gnatp -gnatf},
the wall clock time drops to 293 milliseconds, using 96 threads.

Comparing the 293 milliseconds to the 318 milliseconds with one thread
in quad double precision, we can quadruple the precision
and compute as fast as in quad double precision, using 96 threads,
achieving {\em quality up}.

\section{Conclusions}

Postponing renormalizations of multiple doubles benefits the efficiency.

The convolutions
$\displaystyle \sum_{k=1}^n \sum_{j=0}^i x_{k,j} y_{k,i-j}$
allow to rewrite the inner products
in multiple double arithmetic as matrix multiplications
in double precision floating-point arithmetic,
to prepare for better acceleration with graphics processing units,
in particular tensor cores.

\bibliographystyle{plain}

\end{document}